\documentclass[sigconf]{acmart}

\usepackage{booktabs} 
\usepackage{longtable}
\usepackage{framed}

\acmDOI{xx.xxx/xxx_x}

\acmISBN{978-x-xxxx-xxxx-x/YY/MM. . . }

\acmConference{Conference’17, July 2017, Washington, DC, USA}
\acmYear{2017}

\acmArticle{4}
\acmPrice{15.00}

\begin{document}
\title{Navigating Cultural Diversity: Barriers and Potentials in Multicultural Agile Software Development Teams}

\renewcommand{\shorttitle}{Barriers and Potentials in Multicultural Agile Software Development Teams}





\author{Daniel Welsch}
\affiliation{%
  \institution{University of Applied Sciences and Arts Hannover}
  \department{Dpt. of Business Information Systems}
  \city{Hannover} 
  \country{Germany} 
  \orcid{0009-0007-7439-4750}
}
\email{daniel.welsch@stud.hs-hannover.de}

\author{Luisa Burk}
\affiliation{%
  \institution{University of Applied Sciences and Arts Hannover}
  \department{Dpt. of Business Information Systems}
  \city{Hannover} 
  \country{Germany} 
  \orcid{0009-0006-6784-3346}
}
\email{luisa.burk@stud.hs-hannover.de}

\author{David Mötefindt}
\affiliation{%
  \institution{Agile Move}
  \city{Hannover}
  \country{Germany}
  \orcid{0009-0008-7526-7654}
}
\email{info@agile-move.de}

\author{Michael Neumann}
\orcid{0000-0002-4220-9641}
\affiliation{%
  \institution{University of Applied Sciences and Arts Hannover}
  \department{Dpt. of Business Information Systems}
  \city{Hannover} 
  \country{Germany} 
  \orcid{0000-0002-4220-9641}
}
\email{michael.neumann@hs-hannover.de}

\renewcommand{\shortauthors}{Welsch et al.}

\begin{abstract}
\textit{Context:} Social aspects are of high importance for being successful using agile methods in software development. People are influenced by their cultural imprint, as the underlying cultural values are guiding us in how we think and act. Thus, one may assume that in multicultural agile software development teams, cultural characteristics influence the result in terms of quality of the team work and consequently, the product to be delivered. 
\textit{Objective:} We aim to identify barriers and potentials that may arise in multicultural agile software development teams to provide valuable strategies for both researchers and practitioners faced with barriers or unrealized potentials of cultural diversity. 
\textit{Method:} The study is designed as a single-case study with two units of analysis using a mixed-method design consisting quantitative and qualitative methods. 
\textit{Results:} First, our results suggest that the cultural characteristics at the team level need to be analyzed individually in intercultural teams, Second, we identified key potentials regarding cultural characteristics providing key potentials such as a individual team subculture that fits agile values like open communication. Third, we derived strategies supporting the potentials of cultural diversity in agile software development teams.
\textit{Conclusion:} Our findings show, that a deeper understanding of cultural influences in multicultural agile software development teams is needed. Based on the results, we already prepare future work to validate the results in other industries. 
\end{abstract}

%
%

\ccsdesc[500]{Software and its engineering~Agile software development}

\keywords{Agile methods, culture, diversity, multicultural teams, influence, challenges, barriers, potentials}

\maketitle


\section{Introduction}


Today, many companies are using agile methods and approaches in their software development teams to address the dynamics of the markets~\cite{VersionOne.2022}. This high level of dynamism is manifested in the potential for changing requirements, new technologies, or increasingly shorter product life cycles. To meet these challenges, social skills and cultural aspects such as values and principles are of great importance in agile methods~\cite{Neumann.2023}. While fundamental values and principles are formulated in the Agile Manifesto~\cite{Beck.2001}, we know that the guidelines of widely used agile methods such as Scrum~\cite{Schwaber.2020}, XP~\cite{Beck.2000} or even Lean approaches like Kanban~\cite{Anderson.2011} define further specific values for the application. We also know that social skills, such as communication~\cite{Ram.2019}, are critical success factors. Other success factors include organizational culture~\cite{Claps.2015}, stakeholder integration~\cite{Hoda.2011}, customer collaboration~\cite{Altaf.2019}, and both intra- and inter-team cooperation. 

In recent decades, globalization in the area of software development has increased. We know, that global software development is the term used when cooperation between companies, divisions, or even teams crosses national or continental borders, or even time zones~\cite{Smite.2014}. Also, we know that teams operating in a global software development setting are challenged by various aspects like cultural influences~\cite{Paasivaara.2017} or language barriers~\cite{Ebert.2001}. There are still various motivations and reasons for the globalization in the software development: Purposeful cost reduction, high availability of the teams (e.g., follow the sun principle), or also lack of specialists in the industry nations. We know from various studies, that the effects of global collaboration in software development have been studied in recent years (\cite{CasadoLumbreras.2015,Lous.2017,Smite.2021} to name a few). 

However, cultural diversity in agile software development teams operating in a non-global distributed setting is an increasing phenomenon~\cite{Groschke.2021}. Furthermore, we argue the need for investigation due to the forced recruitment of specialists abroad (from the perspective of the respective company), the cost pressure, and the increasing digitization~\cite{Abadir.2020}. For instance, the Covid-19 pandemic and the comprehensive shift to remote work in many companies have had a further influence on the increased recruitment of experts from other cities and countries as they do not move to the company's site~\cite{Neumann.2022}. Thus, one may assume that cultural diversity will increase in software development teams because the shortage of potential employees with the required skill sets will increase. Especially in the western industrialized nations due to the age situation of the societies
. We know from several studies that cultural diversity may be challenging in software development teams~\cite{Hoffmann.2023}. 

In the context of agile software development, these challenges are highly relevant. For the successful use of agile methods, social skills such as communication, collaboration, or integration of stakeholders are of great importance~\cite{Schoen.2015}. As mentioned above, these social aspects are critical success factors, which is why we need an in-depth understanding of how cultural diversity influences agile software development teams, and what strategies we need to use to address impediments and foster potential. In contrast to global software development, this topic has been underrepresented in literature.

The above motivates the objective of our study, which is refined by the following research questions:
\begin{description}
\item[RQ 1:] Which cultural profiles exist in the teams?
\item[RQ 2:] What cultural barriers or potentials exist in multicultural teams?
\item[RQ 3:] What strategies help to mitigate barriers or to promote potentials?
\end{description}

\begin{table*}
 \caption{Overview of the related work}
  \label{tab1:OverviewofRelWork}
  \begin{tabular}{ccp{0.4\linewidth}}
\hline
Publication & Research design & Findings wrt.\ cultural diversity \\
\hline
Abadir et al. (2019)~\cite{Abadir.2020} & Integrative literature review & The authors found that the intercultural skills for integrating employees with different cultural backgrounds in existing teams is gaining more importance. Covering a wide range of literature dealing with leadership of multicultural in agile organizations the authors identified a research gap. The authors point to the need for further research focusing on the leadership impact on the effectiveness of multicultural teams in agile organizations.\\
\hline
Granow and Asbrock (2021)~\cite{Granow.2021}  & Qualitative study & Based on their results, the authors present a framework indicating that the behaviour of agile teams influencing the performance of multicultural teams. The authors call for future research to get a better understanding how cultural diversity impacts agility on a team level.  \\
\hline
Kohl Silveira and Prikladnicki  (2019)~\cite{Kohl.2019} & Systematic mapping study & The presented systematic maps covers an overview focussing on agile software development. They identified 67 papers dealing with cultural diversity in this area. However, how cultural diversity impacts specific aspects like the software process or the effectiveness of a team was not in scope of the study. Finally, the authors call for research activities combining the several aspects of diversity like age or gender, and cultural background.  \\
\hline
(Preprint) Verwijs and Russo (2023)~\cite{Verwijs.2023}  & Quantitative survey & The authors analyzed how diversity influences the performance of agile software development teams. Their study results show that the cultural background do not contribute to the effectiveness of the teams performance. The authors point to the interesting findings comparing their results with the categorization-elaboration model (CEM). Focussing on the cultural background, the results are not in line with the CEM. They assume that the task interdependence, which may differ significantly among the agile software development teams.\\
\hline
\end{tabular}
\end{table*}

This paper is structured as follows: We give a brief background of the fundamentals of culture and outline the related work in Section~\ref{Sec2:Backg_RelWork}. In Section \ref{Sec3:ResearchDesign}, we explain our mixed-method research design. The results are presented and discussed in Section \ref{Sec4:Results}, including the answers of our research questions. The limitations of our study are described in Section \ref{Sec5:TtV}, before the paper closes with a conclusion and future work in Section \ref{Sec6:ConclusionAndFutureWork}. 

\section{Background \& Related Work}
\label{Sec2:Backg_RelWork}





\subsection{Background}
\label{Background}
According to Hofstede's definition in~\cite{Hofstede.1981}\footnote{Hofstede defined culture as \textit{``The collective programming of the mind that distinguishes the members of one group or category of people from others."}}, we understand culture as the ground for shared values, beliefs, and behaviours of a group or society. Thus, culture guides us in terms of ``what is an expected behaviour" or ``what is implicitly allowed" and 
``what is forbidden". Several authors presented descriptive and comparative models to characterize culture in some ways (e.g., \cite{House.2002,HofstedeMinkov.2013,Karahanna.2006,Trompenaars.2012}). In this study, we focus on the well-known model by Hofstede~\cite{Hofstede.2001} as it is used extensively in the area of empirical software engineering (e.g., ~\cite{Alsanoosy.2019,Ayed.2017,Ghinea.2011}).

\textit{Power Distance Index (PDI):} This represents the extent to which individuals with less power accept or expect unequal distribution of power. An example of this is the way people behave toward their superiors in a company. Societies with a high PDI value believe in hierarchy and take orders without question. Societies with a low PDI strive for equality in the distribution of power.

\textit{Uncertainty Avoidance (UAI):} This dimension describes whether members of a culture feel comfortable or uncomfortable in new or unknown situations. If there is a high level of uncertainty avoidance, the national culture is characterized by clear rules (such as laws or security measures). It tries to create structures that are as clear as possible. There is tolerance for different opinions in national cultures with a low level of uncertainty avoidance. Also, regulations are less precise and strict.

\textit{Individualism vs.\ Collectivism (IDV):} According to Hofstede, national cultures are either individualistic or collectivistic. Low scores are interpreted as collectivist, high scores as individualistic. Members of individualistic societies are interested in their own resources. They tend to care for themselves and their families first and foremost. Collectivist societies share resources and a common understanding of moral standards. 

\textit{Masculinity vs.\ Femininity (MAS):} This dimension describes cultural values in line with traditional gender stereotypes. Societies that score high on the MAS scale prioritize values typically associated with masculinity, such as achieving success, power, and achievement. Cultures with low MAS scores place greater emphasis on relational dynamics and cooperative efforts, which correlate with the IDV dimension.

\textit{Long Term Orientation (LTO):} This dimension is a measure of the extent to which a society values long-term gains over immediate gratification. Long-term oriented cultures often exhibit traits such as thrift and perseverance. Short-term cultures, on the other hand, as exemplified by organizational control mechanisms and executive performance evaluations that focus on short-term results, emphasize immediate results, sometimes at the expense of long-term implications.

\textit{Indulgence vs.\ Restraint (IVR):} This dimension assesses the extent to which a culture permits or restricts the free expression of innate human desires, particularly in the pursuit of life's pleasures. Enjoyment-oriented societies place a high value on leisure time and are less focused on material acquisitions such as rewards.

The model of cultural dimensions by Hofstede has been discussed and criticized in the past decades (e.g., \cite{Baskerville.2003,Schmitz.2014}). Major critic arguments relate to the limitations of few dimensions and that the national level of culture may not be the best unit for cultural characterization~\cite{McSweeney.2002} and pose the risk of stereotyping~\cite{Carmel.2001}. Also, the chosen research design is criticized as surveys may not be an appropriate approach for cultural studies and the representativeness of the population~\cite{Alsanoosy.2020}. Nevertheless, the model is based on a strong empirical basis including a large data set and it has been validated by other researchers. Furthermore, the model applies to several different contexts and has been adopted by various studies (e.g, \cite{Baskerville.2003}). 

\subsection{Related Work}
To provide an overview of the literature related to the effects of cultural diversity in agile software development teams, we searched for both primary and secondary studies. The results of our focused literature search show that the topic has been underrepresented in the recent literature and thus, we identified only three peer-reviewed studies. Thus, we decided to integrate preprints as well and found one more study. To ensure the quality of the preprint, we reviewed the preprint by Verwijs and Russo\cite{Verwijs.2023} considering the Empirical Standards by SIGSOFT~\cite{sigsoft.2023}. The search was performed using Google Scholar because the search engine indexes multiple publishers (e.g., ACM, IEEE). In addition, Google Scholar results show a high degree of consistency with digital libraries such as Scopus~\cite{Yasin.2020}. Table~\ref{tab1:OverviewofRelWork} gives a brief overview of the identified literature.

The research landscape in the field of multicultural agile teams lacks on detail. This is shown by the secondary studies we identified and the calls from other researchers to fill the gap (see Table\ref{tab1:OverviewofRelWork} for details). However, we could identify some interesting findings from other studies. Most of the studies both from the included studies in the literature reviews as well as the primary study by Verwijs and Russo~\cite{Verwijs.2023} focusing strongly on teams performance indicators aiming to understand how diversity impacts the effectiveness using well-known and validated approaches like CEM. Granow and Asbrock present a framework aiming to describe the impact of diversity on teams performance. Interestingly, the findings from both primary studies are contrasting one another in some ways. Verwijs and Russo state that the cultural background does not have a significant influence on the effectiveness of agile teams. However, they also argue that their results may be biased by the task interdependence of agile teams. In contrast, Granow and Asbrock's results indicate such an influence of the individual cultural background. These insights were very valuable for us in designing our research design as we decided to try to create a cultural profile of the teams under study before we analyze the potential impact of the cultural background. 

Although we have searched for related work that deals with challenges and opportunities in culturally diverse agile software development teams, we could not find literature that closely addresses the objective of our study. Hence, to the best of our knowledge, this is the first paper that identifies barriers and potentials in multicultural agile software development teams. 

\section{Research Design}
\label{Sec3:ResearchDesign}


\begin{figure*}[htbp]
\includegraphics[scale=0.21]{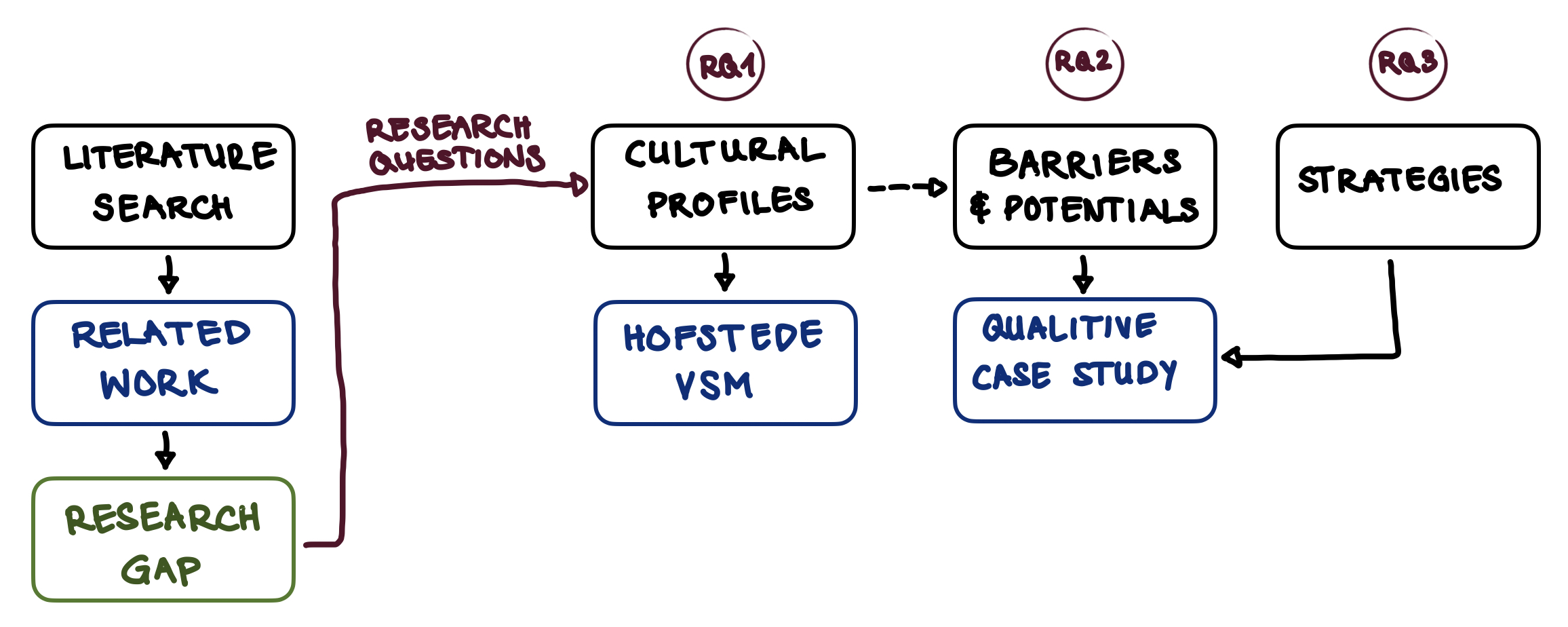}
\caption{Research approach}
\label{fig1}
\end{figure*}

In this section, we explain the mixed-method approach, followed by a brief overview of our research design. We then introduce the case company. Afterwards, we provide a detailed presentation of the research methods used, beginning with the survey by Hofstede. 

The study is designed as a single case study based on the guidelines by Runeson and Hoest \cite{Runeson.2009} using a mixed-method research approach (see Figure~\ref{fig1}). Based on a focused literature review to find studies dealing with topics closely related to our study and the identified research gap, we used the validated survey by Hofstede to be able to analyze and describe the cultural profiles of the teams under study (RQ 1). As we want to identify specific barriers, potentials (RQ 2) and analyze them to provide strategies (RQ 3) to manage and handle them, we decided to collect data using qualitative research methods.

\subsection{Research Context}
We conducted the study at the case company. The research took place at a firm we refer to as Daily Bugle (anonymized), a global enterprise specializing in online marketing with around 1,000 employees.

Our investigation centered on the agile software development divisions within Daily Bugle. The organization operates software development locations across Germany, Poland, the United Kingdom, and the United States. For the purpose of this study, we focused on one of the development sites in Germany. At this site, two agile software development teams are operating, which we refer to as Peter and Ororo for anonymity.

The Peter team consists of seven developers, while the Ororo team has five members. Each team is overseen by a Lead Engineer and adopts Scrum, utilizing two-week sprints. A single Scrum Master assists both teams, and they also receive guidance and coaching on software quality from a dedicated Quality Specialist.

\begin{table}
\centering
\caption{Profiles of the interviewees}
\begin{tabular}{cccp{2.5cm}}
\hline
ID & Current Role & Team & Years working for Daily Bugle \\
\hline
I1 & Agile Coach & Both teams & 15 years \\
I2 & Software Engineer & Peter & 3 years \\
I3 & Lead Engineer & Peter & 7 years \\
I4 & Software Engineer & Peter & 1 year 9 months \\
I5 & Product Owner & Ororo & 2 years 4 months \\
I6 & Lead Engineer & Ororo & 2 years 6 months \\
I7 & Software Engineer & Ororo & 7 months \\
I8 & Software Engineer & Peter & 9 months \\
\hline
\end{tabular}
\label{tab2:Interviewees}
\end{table}

\subsection{VSM by Hofstede}
To ascertain the cultural dimensions of the participants, we decided to use the validated Value Survey Module (VSM) from Hofstede~\cite{Hofstede.2013}. The VSM was administered to the participants through a LimeSurvey questionnaire. This approach was utilized to establish the diverse cultural profiles within the teams Peter and Ororo. To determine the manifestations of cultural dimensions, indices were calculated based on the mean scores of four distinct questions (Details can be found in our research protocol; see Appendix~\ref{Appendix_ResearchProtocol}). 

The survey was conducted over a three-week period in April 2023. We used a dedicated questionnaire in LimeSurvey per team. The respondents rate in team Ororo was 100\% (n=3) and for team Peter 96.5\% (n=6).

\subsection{Qualitative Research Design}
In the case study, eight interviewees, each with distinct roles and varying years of experience in agile software development (ASD), were interviewed (Table~\ref{tab2:Interviewees})

To get an insight into the experience, perception and challenges of participants in the context of cultural variety and collaboration a semi-structured interview \cite{Runeson.2009} was chosen as a methodological approach. This qualitative approach was selected to adjust the discussion sequence of the interview depending on the participants responses. 

With this approach, the case study could be divided into questions about barriers and potentials. To have a systematic method to answer the research questions RQ 2 and RQ 3 we used the Goal-Question-Metric (GQM)~\cite{Basili.2014}. 
Based on the GQM an interview guide (App.D) was created. The guide consists of three parts:
\paragraph{Part 1:} An initial sequence delineating the interview sequences, introducing the interview facilitators, and elucidating the particulars of the case study.

\paragraph{Part 2:} A primary sequence pertaining participants teams. The participant is asked if he would describe his team as international and intercultural. The differentiation serves as an indicator whether a well-established team culture exists within the team. Furthermore the participant is tasked with assessing if their team adhere to the definition of an agile culture ~\cite{Kuchel.2023}.

\paragraph{Part 3:} A concluding section categorizes the interview into inquiries concerning barriers or potentials of multicultural collaboration, alongside neutral queries related to multicultural collaboration. If the participants team adheres to an agile culture ~\cite{Kuchel.2023} they are subjected to targeted inquiries regarding the potentials of intercultural collaboration. If the team does not adhere to it, he is specifically inquired about barriers of multicultural collaboration.

To mitigate potential bias in research outcomes, each interview was conducted by two researchers. One researcher held the interview, the other was observing and taking notes. Moreover, all interviews were recorded and subsequently transcribed to enhance analytical comprehensibility. The interviews and observations were held in May 2023. The interviews were conducted during working hours, not during off-peak hours (at the beginning of the workday, just before the end of the workday). Each interview was conducted using MS Teams and was scheduled to last for 60 minutes. On average, an interview took 18:41 minutes, the shortest interview duration was 14:31 minutes, and the longest interview 23:11 minutes.

We collectively analyzed the transcripts to minimize possible research bias. Each of the transcripts underwent a thorough examination, with an emphasis on identifying barriers, potentials and the strategies already in use to mitigate barriers or to promote potentials. Afterwards, we transferred these findings onto a Miro online-whiteboard (see Appendix~\ref{Appendix_ResearchProtocol}), 
where we analyzed our data using axial coding. Each researcher did the coding by himself to mitigate bias. The results of the coding process were compared afterwards and the differences were discussed in the group. Finally, with an in-depth evaluation of the clustered data, we developed additional strategies to answer RQ 3.

\section{Results}
\label{Sec4:Results}




This section presents the results of our mixed-method study. Initially, we present the cultural profiles derived from Hofstede's VSM to answer RQ 1. Next, we discuss the results derived from the qualitative part of the study to present barriers and potentials of multicultural agile teams and answer RQ 2. Finally, we use the barriers and opportunities we have identified to introduce the strategies developed in response to RQ 3. 

\subsection{Cultural Profiles}
As our study focuses on multicultural characteristics and their influences on agile software development teams, we need to examine the cultural profiles of the team members, because the teams do not consist solely of members with the same cultural background. Thus, in this section, we answer RQ 1: \textit{Which cultural profiles exist in the teams?}

Besides our own data collection using the VSM by Hofstede, we asked for the nationality of the team members. As we were aware of the risk of stereotyping by using cultural characteristics, it is necessary to be able to compare our determined values with them from Hofstede~\cite{HofstedeInsights.2022}. The members of Team Peter are from Germany, Italy and Azerbaijan. The team members working in team Ororo are from India, Pakistan and Turkey. For both teams, the Hofstede data shows different cultural characteristics for the various national cultures. Below, we discuss the determined cultural characteristics at team level comparing them with the values provided by Hofstede. The results are presented for each cultural dimension (see Section~\ref{Background} for a detailed description). We present the specific values per dimension for each team in brackets. However, a figure illustrating this can be found in our research protocol (see Appendix~\ref{Appendix_ResearchProtocol}).

\textbf{Power Distance:} Interestingly, team Ororo (2) has a lower PDI than team Peter (40), although the nationalities of the team members suggest a higher value according to Hofstede~\cite{HofstedeInsights.2022}). Again, according to the Hofstede values, the nationalities in team Peter have very different Power Distance values. However, due to the German team members, a lower value would be expected. This could be a possible barrier for team Peter as one may assume that the team members would not communicate open about their mistakes and failures.

\begin{framed}
\textbf{Answer RQ 1:} First and foremost, the cultural profiles of both teams show differences in half of the cultural dimensions to the values provided by Hofstede~\cite{HofstedeInsights.2022}: Power Distance, Uncertainty Avoidance, and Individualism vs.\ Collectivism. Team Peters cultural profile is collaborative, emphasizes feminin aspects and focus on long term success. The culture also consists of an existing but low Power Distance. The team do not put much importance to uncertainty avoidance. There is a relaxed, casual interaction between team members and their managers. It is important for this team to have fun and enjoyment at work in order to be able to keep their own lives in balance. Team Ororo also values collectivism, but not as much pronounced as team Peter. Otherwise, the power distance do not exist. The team is focusing more on short term success by keeping uncertainty avoidance in mind. In contrast to team Peter, team Ororo values a regulated interaction with each other that focuses on professionalism. Excessive joy, such as laughter, is not welcome and is perceived as unprofessional and disruptive.  The results show, that the teams do have a specific team subculture, which is represented by the cultural profile using the Hofstede dimensions. 
\end{framed}

\textbf{Uncertainty Avoidance:} It is noticeable that team Ororo has a higher level of uncertainty avoidance (35), although three nationalities are represented in team Peter (6), which according to Hofstede have a high value. This could be a potential barrier in team Ororo, as uncertainty and unstructured situations are avoided. 

\textbf{Individualism vs. Collectivism:} Here, team Ororo (35) shows a higher value than team Peter (18). This is interesting, as the national values by Hofstede of the specific teams indicate an opposite distribution of the indices. The low value in team Peter can be seen as a potential for close cooperation and good interaction in groups.

\textbf{Masculinity vs. Femininity:} Upon comparison, it can be seen that the value is higher for team Peter (19) than for team Ororo (1). This is consistent with the values provided by Hofstede~\cite{HofstedeInsights.2022}. According to Hofstede, Germany and Italy in particular showed a high MAS index, which is reflected in the results of our survey.

\textbf{Long Term Orientation:} According to Hofstede, the nationalities in team Ororo tend to have lower LTO, while the nationalities in team Peter tend to have higher one. This can be confirmed in our determined values (Ororo = 33; Peter = 60). The future-oriented actions and willingness of team Peter to adapt represent an important potential. Additionally, the results the dimensions Long Term Orientation and Uncertainty Avoidance confirm each other. Team Peter acts more future-oriented, is willing to adapt and does not avoid uncertainties. Team Ororo is more conservative and avoids uncertainty.

\textbf{Indulgence vs. Restraint:} The nationalities of both teams have rather low indices~\cite{HofstedeInsights.2022}. In the results, it is noticeable that team Peter has a higher value (52) than team Ororo (15). This could be due to the fact that half of the members of team Peter live and work in their home country. The members of team Ororo come from countries where there are large cultural differences, so they may feel more subject to societal values and norms.

\subsection{Barriers and Potentials in Multicultural Teams}
\label{Sec:BarriersandPotentials}
In this section, we discuss our results to answer RQ 2: \textit{What cultural barriers or potentials exist in multicultural teams?}



Openness emerged as a great potential in both teams. 
Different cultures were found to bring new viewpoints and perspectives to the team. Consequently, fostering an environment of respect and tolerance became imperative for effective collaboration among team members from different countries. By cultivating an open minded team culture and promoting transparent communication, the prospects for successful teamwork were enhanced. This is exemplified by one Junior Software Engineer's perspective from team Peter: ``\textit{It requires people to be open for everything and to be willing to meet people and connect with people. And that is what makes this so easy, because all of my colleagues that are from other countries [...] have no issues with talking to you, trying to understand different opinions[...]}".
Similar opinions were also expressed in the interviews of the team Ororo.

During our observations of team Ororo, we noticed that a lack of open communication disrupted teamwork, and extended pauses between conversations could be identified.
This phenomenon may lead to unaddressed issues within the team, potentially causing team members to experience heightened levels of stress and pressure.

Team Peter characterized their communication within the team as very open and direct. This fosters a culture of mutual assistance, constructive feedback and agile practices such as pair programming.

The openness influences the personal relationship between the team members. The participants of both teams expressed a positive relationship with their colleagues and an interest in meetings outside of work.
Teams which meet outside of work tend to have a better team performance.

After a team has been working together for a while, a team culture (subculture) can grow
. The more established this subculture is, the better the team's internal cooperation works 
. This subculture can mean that a collaborative team customizes the agile approach.
A Software Engineer of Peter said: ``\textit{We just tailor Agile to our needs, which is in the end what I believe is being Agile, because everyone has different requirements as a team}".

The phenomenon of a distinct team could be noticed in the observations at agile events. Direct and respectful communication within the team with a subculture was observed, with criticism being expressed openly.

The team with the less distinctive subcultures showed restraint when questions or ambiguities arose. It was noticeable that all opinions were accepted and there was no discussion. The stand-up did not follow the agile practice, instead the focus was on the backlog.

A discrepancy was found in Team Ororo between the portrayal of direct and open communication in the interviews and the observations. The results of the questionnaire show a high UAI in the Team Ororo. Reticence among the team members in the agile practices could be observed. The reason for this may be the team composition, which has only existed for a few months
. Team Ororo has not existed long enough for a subculture to develop and establish itself. The prevailing impression suggests that team members are prioritizing completing individual tasks, often at the expense of teamwork.

Teams with a history of collaboration exhibit a pronounced team dynamic. This was particularly evident during the observations. They cultivate a distinctive subculture, which can lead to new potentials. A pronounced collaboration can be confirmed by the interviews and the questionnaire. The Individualism Index is low in teams with a subculture because the team members are adaptable.

\begin{framed}
\textbf{Answer RQ 2:} A team culture that emphasize openness represents a key potential of cultural diversity in multicultural teams. It provides the opportunity for new viewpoints and perspectives. Furthermore it supports Kaizen and thus, is eminent for the success to create an agile culture. In turn, teams which do not put focus on creating its own subculture are impeded by hindered feedback, criticism and thus, identifying optimization potential. 
\end{framed}

Another potential is the individualization of the Agile Manifesto to the needs and demands of the individual multicultural team. 



\subsection{Strategies to manage Barriers and Potentials}
The following section outlines strategies for addressing RQ 3: \textit{What strategies help to mitigate barriers or to promote potentials?}

A strategy arises from team members' personal meetings outside of work. This fosters the development of a distinct subculture in the team. Therefore, successful teams should continue these personal meetings to preserve and reinforce the potential. In addition to leisure-time gatherings, team-building measures within the teams can bolster trust and alleviate uncertainties.

Furthermore, we found reinforcing a feedback-culture and integrating into the daily schedule can faciliate a shared understanding of feedback and account for cultural differences. In this regard, a guideline or regulatory framework can be supportive in its formulation.

The high workload suggests that team members lack the time for team-building activities, as a result, a team culture cannot develop. One of the interviewees mentioned that not even a 15-minute coffee break, scheduled every two weeks, could be adhered to. 
One strategy is to give new teams or team members a lighter workload initially to help them integrate into the multicultural team. To support this integration, scheduled coffee breaks for personal interaction should be implemented more frequently.

Another hypothesis is that as the age of a team increases, its performance capabilities also increase. This is attributed to the well-established subculture within a team. Therefore, organizations should prioritize long-term employee retention as a means to sustain the team in the long run. Special emphasis should be placed on multicultural backgrounds.

Our identified potentials and barriers (see Section~\ref{Sec:BarriersandPotentials}) by creating such a subculture is confirmed by the team members understanding of what an agile culture should be. The team members from both teams agree with the definition by Kuchel et al.~\cite{Kuchel.2023}: ``\textit{An Agile Culture reflects to the behaviour of people working in an organization using agile practices based on the underlying values and principles defined in the agile manifesto and the guidelines of agile methods.}"

Thus, we finally recommend to be aware of the importance of value-based work emphasizing the specific cultural characteristics, which supporting important aspects like open communication or feedback. However, as mentioned above, such a development of a teams subculture that fit with the attributes of an agile culture needs time and patience. 

\section{Threats to Validity}
\label{Sec5:TtV}



Despite adhering to established guidelines for designing our single case study and for the data collection and analysis process, it's important to take some inherent limitations into account.

\textit{Construct validity:} Designing the interview guideline, we applied the Goal-Question-Metric (GQM) methodology, a well-validated and widely-recognized approach. To ensure quality, the document was initially formulated by a research team of five members and subsequently reviewed by an agile software development expert and one senior researcher. Our interview guide consists of both, closed and open-ended questions to obtain targeted insights. Additionally, the questions were designed to be non-leading to minimize influence on the respondents from the agile software development teams. We also provided an option for respondents to add their own answers when the preset choices were not applicable.

However, the quantitative nature of the survey limits the depth of understanding we can achieve, especially when presented with intriguing or diverse responses. There's also the potential for bias among respondents due to their past experiences. For example, the teams in the study have been working remotely for over two years, which could lead to a focus on the positive aspects of their experience. 

\textit{Internal validity:} Although we performed a thorough analysis of existing literature, some internal validity threats are worth to mention. It is a significant challenge to ensure that we identified the relevant literature. Based on a systematic approach, we tried to find and analyze as much relevant work related to our study as possible. Thus, we decided to search in two digital libraries (Google Scholar and Scopus). As the survey by Hofstede is validated several times, we add another layer of robustness to our research approach. 

\textit{External validity:} The study's external validity is constrained due to its focus on a single case study. To enhance the generalizability of the findings, future research could include additional cases from not only other software development departments within Daily Bugle but also from companies in diverse industries and countries. Nevertheless, we aimed to minimize other variables that could affect the outcomes, such as the development stage of the agile teams being studied or the unique circumstances surrounding product development. It's worth noting that the response rate for our single-case study was high and encompassed both agile development teams as well as the Scrum Master. As a result, we are confident that the data collected provides a meaningful representation of the situation in this specific case.

\section{Conclusion and Future Work}
\label{Sec6:ConclusionAndFutureWork}



This paper presents the findings of our mixed method study dealing with the potentials and barriers of intercultural agile software development teams. Additionally, we present strategies to promote the potential and mitigate the barriers of cultural diversity based on a discussion of our results. 

We identified that a team subculture, which emphasized agile values is a great potential for intercultural teams, as it provides the opportunity for open communication, easier integration of new colleagues and it fosters new viewpoints on specific problems or challenges. These aspects are eminent for the success of agile software development teams. 

However, creating a subculture needs time and patience from both the team members as well as their stakeholders and organization around them. Thus, we analyzed our results to create specific strategies promoting the potential and mitigating the barriers of intercultural agile software development teams. We recommend to focus on social activities to encourage the team members for taking their time to integrate new colleagues, discuss problems and find solutions together. Furthermore, team events may be a good idea to foster a team's subculture. These team events could be coffee breaks, team breakfast sessions, or even team building events outside the remote or onsite office. Also, it is important that such a subculture relies on the team's composition. Thus, we recommend to focusing on the employee's well-being and foster long term employee loyalty and create a strategy to decrease the turnover of the team members.

Our findings support both practitioners and researchers agile community as one may assume that cultural diversity will increase in future. However, as we designed the study as a single case study, we call other researchers to investigate this phenomenon in other industries or contexts (like regions). 

\appendix
\section{Appendices}
\subsection{Research Protocol}
\label{Appendix_ResearchProtocol}
As recommended by Runeson and Hoest~\cite{Runeson.2009}, we created a research protocol for documenting our research activities. The research protocol is available at \href{https://figshare.com/s/1d9460813cbe49b36f62}{FigShare}.

\begin{acks}
The authors want to express their deep gratitude to the participants to the study and the support from our case company Daily Bugle. 
\end{acks}

\bibliographystyle{ACM-Reference-Format}
\bibliography{references} 

\end{document}